# Digital Limits of Government: The Failure of E-Democracy [*]

Zach Bastick
Department of Government, Harvard University, Cambridge, MA, USA
Email: zach.bastick@gmail.com

## Abstract

While the Internet is often touted as a revolutionary technology, it might be noted that democratic institutions have witnessed no digital revolution through the Internet. This observation leads this chapter to argue that the field of e-democracy has generally failed to live up to its own reformist rhetoric. It argues that instead of reforming government processes through technology, e-democracy projects have tended to focus either on lowering the costs and increasing the efficiency of existing political processes or on analysing the civic participation that occurs outside of purpose-built e-democracy platforms. The chapter suggests that this lack of attention to the Internet's potential for systemic change in formal political institutions has little normative impact on the democratization of society and may even re-enforce, rather than challenge, the sociopolitical status quo. Further, it suggests that the current approach of e-democracy risks normalizing the Internet to the norms and expectations of the offline world. To elucidate this argument, this chapter overviews both the general trend of e-democracy projects and criticisms of those projects. Finally, the chapter proposes a more radical vision of e-democracy that, it suggests, would usher a larger potential for democratization. This more radical vision of e-democracy consists of recognizing the attributes of the Internet that transcend the limits of the analogue world and applying these to democracy. Such an approach would open the path for envisaging new political processes and systems, allowing the field of e-democracy to live up to its own rhetoric, and affording society the means to address multiple of the centuries-old problems faced by democracy.

---

[*] *Cite this chapter:*

## Introduction

The relatively new fields of e-government and e-democracy have largely focused on the potentials of the Internet to support and deepen democracy. They have sought to leverage the Internet to promote more transparent government, increased accountability, citizen-centred services, distributed association, simplified petition and contestation, more responsive representatives, and the re-engagement of citizens. Yet, while these advances introduce technology into government, they fail to leverage technology to fundamentally transform government. This chapter will discuss the limitations of e-democracy projects in defining a strong direction for reforming government through technology. It will build the case that the potential of e-democracy touted by the discipline's own rhetoric has not been achieved and that, to more fully achieve this potential, a radical vision of e-democracy must be adopted. To develop and support this argument, the chapter will, first, situate reformist rhetoric within the early radical aspirations associated with e-democracy. It will then argue that both e-democracy projects and criticisms of those projects have largely diverged from these aspirations—and that this divergence may have contributed to promoting incremental improvements in government processes over fundamental advancement of those processes. Finally, the chapter will argue that future research needs to move beyond the framework of past approaches in order to fulfil e-democracy's core promise of fundamentally reforming government and deepening democracy.

## The Lost Foundations of E-Democracy

The aspiration to democratize society through electronics can be traced to an epoch well before the Internet, and indeed before the mass commercialization of personal computers—e-democracy might be said to have arisen from an electrical vision of distributed democracy in a largely analogue era. On 9 April 1940, Buckminster Fuller, the American author and architect, envisioned a system of futuristic distributed voting whereby citizens would vote from their homes using telephones. Following the commercialization of television, in the 1962 preface to his book, Fuller extended this possibility to two-way televisions. At a time of growing dictatorships, Fuller wrote that electrified voting '…promises a household efficiency superior to any government of record because it incorporates not only the speed of decision which is the greatest strength of the dictator, but additional advantages which can never be his' (Fuller 1971, p. 11). Having citizens vote on issues from home would be collaborative—certifying 'spontaneous popular co-operation in the carrying out of each decision' (Fuller 1971, p. 11). This constant input from citizens would allow a rational progression of the governance process—'continuous correction of the course, or even complete reaction, should (and as) experience indicate desirability, without political scapegoating' (Fuller 1971, p. 11). In telephones and television, and more precisely in their architecture, Fuller saw the potential for a more active role for citizens, as well as the establishment of a new relation between the citizen and the state.

The e-democracy literature rarely explicitly recognizes its lineage to Fuller. Yet, the significance of a tracing e-democracy to Fuller is not only historiographical—it is also thematic and methodological. While chronology might place Fuller's proposal on the outskirts of what might contemporarily be considered as 'e-democracy', the concept of household voting envisioned by Fuller can be situated within the e-democracy lineage because it reassesses political processes through information and communication technologies (ICTs). In this sense, Fuller's 'electrified voting' transcends simple novelty in engineering to breach the realm of technology with the realms of political theory and philosophy. He presented no plan for how such a system would work in practice but proposed its possibility—the system may work or it might not work, but, he argued, it should at least be considered (Fuller 1971, p. 12). The qualities of this consideration, even if intangible and practically unrealized, provide a bedrock for what radical e-democracy might become, even if it still has not almost a hundred years later: it challenges the perceived limits of the physical world, highlights the ideals of our societies, and reconsiders how democracy might attain these ideals.

Political theory and philosophy have played a central role in the development of e-democracy, even if these considerations might be minimized today in favour of a focus on the incremental development of government through technology. The distributed, neutral, and fast nature of the Internet distinguishes it from previous large-scale communication technologies, including from those Fuller toyed with, such as radio and television. The architectural uniqueness of the Internet has been cast, especially by academics, as an opportunity to democratize society to unprecedented levels. Perhaps this is in part because the digital world allows bypassing constraints on democracy imposed by the non-digital world. The 1990s saw the growth of a technological counterculture, with the growth of utopian rhetoric predicting an 'Internet revolution'; indeed, even politicians, such as Howard Dean, explored the prospect of democratic deliberation through electronic town meetings. In the 2000s, the collaborative and distributed nature of the Internet leveraged by Web 2.0 highlighted the potential for participatory renewal through collaborative technology. In the 2010s, the use of technology by social movements around the world to associate, challenge the status quo, and sometimes topple totalitarian governments from the ground up further fuelled the rhetoric associated with e-democracy. Paulin (2015) has written of 'powerful myths' fuelling the integration of technology and government—and indeed, the framing of e-democracy as a revolutionary field has been shaped by the amorphous and varied phenomena, platforms and events of the last three decades.

Yet, despite the rhetoric, it seems implausible to attribute fundamental political changes to the field of e-democracy. Normatively, the role of political theory and philosophy is perhaps minimized today in favour of playing 'catch up' with the technological developments. Empirically, e-democracy has focused either on (1) building platforms for democracy or (2) analysing use of the Internet occurring outside of e-democracy platforms. E-democracy platforms include e-voting (the online version of voting), e-petitions (the online version of petitions), and portals that allow citizens to email their representatives (the online version of telephoning or writing a letter to government). These tend to leverage the Internet's architecture to lower the costs of existing processes or to make these processes more efficient. Yet they have for the most part failed to galvanize public opinion, foster a mass renewal in participation, challenge core problems of democracy, or fundamentally reform political processes.

On the other side of the e-democracy field are projects that apply outside theories and constructs (especially social, psychological, and economic constructs) to analyse participation occurring on the Internet that is perceived as being significant to democracy. For example, a multitude of e-democracy studies have analysed social media that have facilitated the decentralized organization of mass movements (such as Occupy) as well as revolutions (such as the Arab Spring) through the Internet. While both the platforms developed by e-democracy and the analysis of participatory politics occurring outside of e-democracy borrow and add momentum to the radial rhetoric of e-democracy as a field, it is rare that these two approaches to e-democracy fundamentally question the democratic processes or envision the creation of new ones through technology. Further, it is perhaps telling that the activity deemed to have democratic significance that is most frequently analysed in e-democracy is activity occurring outside of purpose-built e-democracy platforms.

The attention that the Internet has drawn for its potential for radical change appears not to have resulted in any radical change for democracy: power relations within Western societies have perhaps been entrenched rather than altered, and the centuries-old structures that in democracy are expected to translate societal values to political outcomes remain largely unchanged. The field of e-democracy which by its name asserts to bridge electronic technology and democracy, and further to this might be assumed to promote democracy, has been largely inert in attempting to realize radical changes to democracy. To grasp the extent of this inertia, the following section will explore the approach of e-democracy towards 'civic participation', which due in part to its central significance for political theory, is a common theme in e-democracy research.

## Wariness Towards Digital Novelty: The Example of Participation

Radical visions of e-democracy have met strong criticism from within the e-democracy field that challenge the significance of Internet phenomena to formal political application. It was a common aspiration amongst the utopians of the 1990s that the Internet would democratize society. This was largely associated with the Internet having a distributed architecture and allowing cheap and neutral communication over large distances. It was also coupled with high expectations of citizens' motivation, ability, and wishes—framing the Internet as a conduit for such democratic virtues as increased citizen participation in government. Hence, multiple models of e-democracy have taken the approach that providing a platform for participation through the Internet will itself contribute to democratizing society. This 'if you build it they will come' was based on, as it has been framed by Paulin, a 'purified image of a reinvented, new and better government, serving a new generation of omnirational, tech-savvy citizens' (Paulin 2015, p. 1). While some scholars have theorized that the Internet does promote engagement because of its architecture and the cyberspace that emerges from it (e.g. Benkler 2006), numerous academics have argued against this, suggesting that the Internet either does not provide significant opportunity of participation or that it does not stimulate political engagement. Indeed, amongst others, Buchstein (1997), Sunstein (2009), Margolis and Moreno-Riaño (2013), and Shapiro (1999) have argued that Internet use may weaken rather than strengthen political participation. Common amongst these views is that, 'central features of the Internet… generally undermine the sort of public sphere and political interaction that is required for genuine democratic deliberation' (Bohman 2004, p. 131), thus fundamentally limiting Internet-based e-democracy. These two criticisms—that the Internet does not provide the conditions necessary for democracy and that e-democracy overlies on an unreasonable expectation of participation—are two prevalent criticisms of the field that may have stumped the development of Internet-based participatory democracy models.

The premise of these criticisms is partly based on the notion that online exchange lacks some of the communicative intricacies of face-to-face interactions. For one current of thought, this notion promotes equality and rational discussion—the minimized social context of Internet-based communication may bode well for democracy by reducing socio-economic prejudice as income, education, race, and skin colour are not immediately displayed (Witschge 2004, pp. 116–117). Yet for another current, computer-mediated communication does not eliminate such discrimination but supports the development of non-visual methods of identifying socio-economic qualities and alternative criteria for judging others (Kollock and Smith 2002). Further, beyond the immediate lack of general socio-economic cues, there may also be a lack of the social cues that are postulated to foster thick online communities, including those indicating trust, familiarity, stability, and social pressure (Van den Hoven 2005, p. 53). The lack of these hints, it is argued, reduces social interaction to a state where association is impeded, thus 'eroding the supply of social capital so critical to democracy' (Chambers et al. 2005, p. 125), and consequently limiting the formation of cohesive communities that support genuine, rational–critical exchange. For others, the online world is so far removed from what is perceived to be the 'real world' outside of the Internet that heavily relying on the Internet threatens awareness of reality. According to this perspective, communicating extensively through computers leads to a process of 'de-realization', which in turn creates artificial desires and needs (Buchstein 1997, p. 250). By distancing citizens from political and social realities, computer-mediated communication may limit the ability of citizens to ascertain issues and grasp the viability of potential solutions. The general notion underlying these criticisms of e-democracy is that the Internet and ICTs are unable to provide as rich of an environment as the offline world, and it is consequently concluded that any form of e-democracy that relies heavily on online association and deliberation will necessarily be less effective than the current, non-virtual democratic processes of the offline world.

A similar criticism of e-democracy is that the quality of democratic participation on the Internet is not sufficient to fulfil the assumptions of many normative models of democracy and may instead lead to political polarization. From one perspective, this has to do with the lack of inclusion in the Internet demographic. From another perspective, it has to do with the dynamics of online participation and the outcomes of that participation. Anonymity and distance promoted by computer-mediated communication have been postulated to encourage uncivil deliberation that is neither rational nor critical. This finds evidence in discourteous, unaccepting, and disrespectful online behaviour. The phenomenon of 'flaming' is perhaps the most obvious of these. Although Witschge recognizes that 'anonymity and the absence of social presence, which seemed so promising for democracy, can instead work against a genuine democratic exchange' (Witschge 2004, p. 115), she advocates that rules and guidelines can increase civility and notes further that 'we do not know whether these online uncivil behaviours have the same effect as the offline ones would have' (Witschge 2004, p. 116). Additionally, there is a fear that Internet users are unwilling to critically question their own opinions and weigh those of others, but instead seek out interaction with like-minded individuals (Bellamy and Raab 1999; Sunstein 2009). According to Bellamy and Raab (1999), this would lead to a balkanization of politics and a fragmentation of the online sphere. Such views have been largely supported by empirical studies of online deliberation (Wright and Street 2007, p. 852). Buchstein (1997) assimilated this to a process of 'show and tell' wherein users exposit their arguments independently from others, but doing so in such an aggressive manner that it becomes a process of 'show and yell'. Finally, there is a fear that if electronic discussions gain importance, it would foster extreme views as individuals are increasingly fostered within their own communities. This strongly challenges the ability of the Internet to inform citizens and form consensus.

An additional common criticism of e-democracy is that it overly relies on rational participation. Whether online or offline, and whether through deliberation or voting, participatory democracy demands political engagement from citizens. Citizens are expected to take time to understand, review or monitor, and act on the political agenda—a far cry from the sometimes-held idea of the modern citizen as alienated and apathetic. Technology can be viewed as either enhancing or reducing political various ideals. As Iyengar and Ansolabehere (2010) argue, for instance, campaign advertising over televisions help to inform citizens about candidates, but may promote a 'spectacle democracy' that supports consuming politics rather than actively contributing to it. Support that new approaches to technology might (contrary to traditional uses of television) promote participation came perhaps most fervently from Becker and Slaton, who claimed that Televote projects 'profoundly contradict the portrayal of the American "couch potato", a nation full of potbellied male dolts who would never swap their six-packs of beer and recliners in front of the television set for Styrofoam cups and plastic chairs at a public policy forum' (Becker and Slaton 2000, p. 95) and insisted that 'almost all of the designers of these projects have been amazed at the gratitude of ordinary citizens for being asked and included' (Becker and Slaton 2000, p. 95). Yet, criticism of this basic premise that citizens are motivated to participate more actively in politics, a criticism to which Becker and Slaton were responding, has been an explicit challenge to e-democracy from the earliest days of the formal discipline. For instance, as early as 1987, in his survey of television and telephone-based e-democracy, Christopher Arterton 'found little support for the notion that citizens have the interest necessary to sustain near universal participation' (Arterton 1987, p. 197). Still today, along these lines, e-democracy is sometimes viewed as being unrealistic and overly ideal—as Matt Qvortrup states, e-democracy is seen as 'an ideal pursued by super-engaged citizens, not as a serious contribution to increasing public participation' (Qvortrup 2007, p. 67). In addition to the question of whether citizens are motivated in the absolute sense—that is whether they would participate more directly in politics if given the chance—is the question of whether citizens may be motivated by ICTs in the relative sense—that is whether the Internet might foster greater participation than did the pre-Internet world.

Despite optimism from utopians that the Internet will lower boundaries to participation, the validity of the expectation that the Internet will engage users has been the focus of both theoretical and empirical challenges. One approach is that the sheer amount of information online increases information costs above any reduction that might occur because of the Internet's architecture. This argument that users suffer from 'information overload', whilst being of a rationalist perspective, suggests that the Internet may actually be detrimental to democracy as too much information is spent on filtering information rather than accessing new information. As van den Hoven states, 'the average citizen is not willing to incur the information cost and transaction cost associated with political deliberation' (Van den Hoven 2005, p. 53). Another approach is that the rationalist perspective is too narrow and that a simple cost-and-benefit perspective fails to address more complex problems inhibiting motivation. Christopher Arteton, writing about television and telephone models, postulated that lowered costs of communication would not increase participation to the degree required by plebiscitary notions of e-democracy. For Arteton, the issue of motivation appears to transcend rationalist weighing to a deeper-rooted lack of interest in politics—he writes that 'in practice, too few are interested enough to make plebiscites a feasible means of policy making' (Arterton 1987, p. 197) and estimates that if citizens were given the opportunity for direct democracy through ICTs, 'probably around two thirds will not participate' (Arterton 1987, p. 197). Although Bimber (1998) suggested that the Internet may allow for higher paced government in which representatives respond to issues from emergent groups, he also dispelled the ideal that the Internet increases participation from the bottom-up. Bimber noted that although cross-sectional data does show that citizens who are more informed participate more than those who are less informed, longitudinal data does not support this. As he writes, 'none of the major developments in communication in the Twentieth Century produced any aggregate gain in citizen participation. Neither telephones, radio, nor television exerted a net positive effect on participation, despite the fact that they apparently reduced information costs and improved citizens' access to information' (Bimber 1998, p. 57). Turning to the Internet, Bimber found that political interest was less associated with using the Internet than watching television or reading newspapers, and that donating money was the only type of participation that aligned with Internet use (Bimber 2001, p. 53).

## New Technology, Old Ideas: The Core Problem and Risks of E-Democracy

What is missed by the usual conversations in e-democracy, but becomes increasingly obvious as the field's research is contrasted with its rhetoric, is that new technology permits the exploration of new ideas. Having largely circumvented this point, e-democracy has replicated old ideas using new technology. This might be at least in part due to a lack of awareness as to what makes the Internet unique in relation to the pre-Internet world, or to a lack of creativity as to how to apply this uniqueness of the Internet to politics. The dangers of this tendency of e-democracy to apply the Internet to further the status quo are multiple. First, there is missed opportunity of unrealized potential. Second, there is the risk of normalizing the Internet to the offline world through future technical, social, or legal decisions—chopping down the unique tangled branches of the Internet so as to form a ubiquitous, but standardized and characterless square that more easily integrates our current development needs.

The Internet has largely been applied to further the political status quo rather than exploring alternative democratic futures. That is instead of realizing the potential of the Internet to produce alternative power structures or political processes, e-democracy has largely sought to assess how the Internet can replicate existing democratic processes: Bulletin Board Systems, Usenet, Facebook, Twitter, and Meetup may have facilitated association and civic action at previously unattainable levels, but when used in e-democracy they largely reproduce the processes behind or provide an interface to, offline association and communication; urban democracy projects such as Paris' DansMaRue, which allows Parisians to report broken benches or graffiti to the city, do little more than replace other municipal communication channels; local democracy platforms, such as the use of Twitter in the Spanish town of Jun, supplement or replace bureaucracy, but keep fundamental processes of representation in place; online campaigning and new media may provide an edge to tech-savvy campaigns—facilitating political communication in much the same was as they did for Howard Dean, or increasing the base of campaign financing as they did for Barack Obama—but they do not deeply change political systems; e-voting and online petitions may facilitate consultation of citizens by governments but, even in nations such as Estonia that have deployed electronic voting at the large scale over multiple election cycles, these new tools of e-democracy only reproduce existing processes behind offline voting and contacting representatives.

This is not to say that e-democracy exactly replicates offline 'tools' of democracy—online petitions on an official government website, with supporting legislation, do facilitate identifying policy issues and public opinion distinctly and more efficiently than do form letters or phone calls to representatives, and likewise, online voting distinguishes itself from offline voting, and through this distinction might lower burdens and increase participation. However, it is to say that for a large part e-democracy reproduces the 'processes' behind offline forms of participation. For example, online voting automates, but does not change the way electoral systems work—candidates must still be presented to citizens, citizens must still actively cast votes, and votes are still tallied similarly to how they were before. Indeed, by the time an election ends—and citizens have streamed campaign videos to their mobile phones, commented on attack ads in threaded discussions, associated through social media groups and hashtags, assessed and discussed the political platforms of the candidates, and optimistically clicked their mouse buttons on the name of their preferred candidate—it may very well be that, in the grand scheme of politics, representation, and power relations, very little has changed.

Similarly, criticisms of e-democracy fall into the same trap of replication—but replicate assumptions rather than processes. For instance, the retort that e-democracy should avoid deliberate models because the Internet does not uphold many assumptions of the offline world sidesteps the potential for deliberate models based on alternative assumptions. Whereas it is conceivable that the Internet might not provide the social cues necessary to foster the 'thick communities' of the local town hall, or the 'rational deliberation' of the highly formalized meetings of experts and bearcats, it is also conceivable that it provides other assets for democracy. Indeed, e-democracy could frame the uniqueness of the Internet as an asset for democracy—and through this reframing, thick communities consisting of anchored identities may give way to ephemeral communities of fluid identities, wherein membership may be dynamic in much the same way that human spontaneity and intellectual exploration can be dynamic. Similarly, anonymity might be reframed from permitting unconstructive discourse (e.g. flaming and trolling) to, instead, allowing the removal of an identity-centred threat imposed on the civic discourse of individuals by society. The lack of the threat of having one's intellectual production linked to a single personal identifier (such as a name) might free individuals from society's traditional reliance on consistent identities, to allow individuals to participate with alternative and even conflicting identities. If e-democracy recognizes these unique attributes of the Internet, it can explore a larger array of alternative political futures.

It is unclear as to whether the prescriptive approach common in e-democracy will continue and, if it does, what the relevance of the field will be to modernity. In a very specific sense, as Internet penetration has increased and user behaviour shifted, it is questionable whether many of the criticisms mentioned in the last section remain true today. For example, contrary to Bimber, a cross-modal study by Ansolabehere and Schaffner (2014) found that, on three nearly identical surveys conducted by Internet, telephone and mail, opt-in Internet, respondents were more politically informed than respondents from the other modes. In addition, they were more likely to report contributing to political campaigns and to obtain less of their news from television (Ansolabehere and Schaffner 2014, pp. 11–12). Xenos and Moy (2007) found from an analysis of the 2004 American National Election Studies that there is only mixed support for both the rationalist and the psychological approach—the authors found that information acquisition and use conforms more to the rational perspective while more general political and civic participation conforms more to psychological perspective. More generally, the simultaneous development of Internet phenomena and the integration of the Internet in daily life increasingly highlight the tensions between ageing political institutions and the modern society that those institutions serve. From this more general perspective of the zeitgeist, it is conceivable that there might be an increasingly evident societal need for a reformed e-democracy field that is Internet-sensitive and non-prescriptive.

The dangers of continuing the current approach of the e-democracy field are multiple. First, framing the digital world through traditional, pre-digital perspectives limits advances to those that fit within those traditional perspectives. This may suppress the emergence of new perspectives. Like Maslow's law of the instrument—'if all you have is a hammer, everything looks like a nail' (Maslow 1965, p. 15)—viewing the digital world through a pre-digital lens obscures new aspects of the Internet that may be fundamentally different from the offline world, and thus whose application might fundamentally impact the offline world. In the field of e-democracy, seeking out how Internet technologies can be mobilized within the confines of our existing governance risks maintaining the status quo and overlooking alternative political structures and futures. This favours incremental changes based on the lowest common denominator of shared attributes between the Internet and society. This is in part because applying the architecture of the Internet to existing political structures may have quantitative results (such as increasing efficiency and lowering costs), but does not bring into question the underlying processes that those structures support. Second, disregarding the unique attributes of the Internet may in turn minimize or suppress these attributes in a process of normalization. Without a base to protect these unique attributes of the Internet, the Internet may be normalized to the offline world. This may occur to fit the traditional expectations or structures of society. Over the last two decades, this has been witnessed in digital access and controls aiming to protect pre-digital commercial interests (such as the movie or music industries) from emerging methods of association and cultural production (e.g. peer-to-peer and remix culture). Finally, the incremental developments favoured by e-democracy, taken together, risk further strengthen the injustices and structural weaknesses of democracy. Indeed, by shaping technological advances after existing institutions, e-democracy may act to consolidate the ascendency of existing institutions over their potential alternatives—leading some to observe, for instance, that 'the digital era seems to be merely another noteworthy change in environment which the bureaucracy aims to survive' (Paulin 2015, p. 4). This strengthening may serve to extend the life of political institutions or processes that would otherwise be replaced or have a greater need to be replaced, in line with the continual evolution of societal values. In short, being more sensitive to how the unique aspects of the Internet can fundamentally improve governance would both encourage developing alternative perspectives for democracy, protect the organic evolution of the Internet, and promote democratic renewal.

## Conclusion

The tragedy of e-democracy is that the field misses the opportunity to fulfil its own aims. This chapter has argued that while the Internet presents a myriad of opportunities for radically democratizing government, the core field expected to explore democratization through technology has largely sidestepped these opportunities. The role of technology in affecting power relations

Page **12** of **14**

and control structures appears to be increasingly subordinated to the political status quo. E-democracy has focused on incremental developments to democracy, many of which consist in applying the efficiency and cost reduction enabled by the Internet to streamline or replicate, albeit in digital form, existing government processes. Hence, the participatory tools proposed by e-democracy strongly resemble those of the offline world, and the approach of modern e-democracy can be situated well before the popularization of the Web and the emergence of Web-oriented concepts. E-democracy's apparent outcomes of increased efficiency, greater inclusion of marginalized groups, and facilitated access to information have been incremental rather than structural—and consequently no 'Internet revolution' has occurred in government. By applying new technology to political structures that are centuries old, not only is e-democracy missing a large opportunity for impacting democracy through technology, but it also risks normalizing the Internet to the norms of the offline world, as well as emphasizing the current political system, with its structural vulnerabilities and the perceived injustices of its political outcomes. In conclusion, e-democracy must transcend its current ideology of incremental change in order to tap into a larger potential for fundamental change. New directions, including Internet Democracy, liquid democracy, peer-to-peer governance, block chain democracy, decentralized autonomous organizations, and wiki-based government are all examples of future directions that e-democracy might take in order to move digital-era government beyond the vestiges of the industrial era—that is 'beyond bureaucracy'.